\documentclass{jltp}

\usepackage[english]{babel}
\usepackage[latin1]{inputenc}
\usepackage[dvips]{graphicx}
\usepackage{bm}
\usepackage{amsfonts}
\usepackage{amssymb}
\usepackage{amsmath}

\newcommand{\equref}[1]{(\ref{#1})}

\title{Detecting Current Noise with a Josephson Junction in the Macroscopic Quantum Tunneling Regime}

\author{J.T. Peltonen$^1$, A.V. Timofeev$^{1,2}$, M. Meschke$^1$, and J.P. Pekola$^1$}

\address{$^1$Low Temperature Laboratory, Helsinki University of Technology,\\ P.O. Box 3500, 02015 TKK, Finland\\
$^2$Institute of Solid State Physics, Russian Academy of Sciences,\\
Chernogolovka, 142432, Russia}

\runninghead{J.T. Peltonen \textit{et al.}}{Detecting Current Noise
with a Josephson Junction}

\begin{document}

\maketitle

\begin{abstract}
We discuss the use of a hysteretic Josephson junction to detect
current fluctuations with frequencies below the plasma frequency of
the junction. These adiabatic fluctuations are probed by switching
measurements observing the noise-affected average rate of
macroscopic quantum tunneling of the detector junction out of its
zero-voltage state. In a proposed experimental scheme, frequencies
of the noise are limited by an on-chip filtering circuit. The third
cumulant of current fluctuations at the detector is related to an
asymmetry of the switching rates.

\end{abstract}

\section{INTRODUCTION}
Complete description of charge transport through a mesoscopic
conductor requires knowledge of the probability distribution of
current through the conductor. In general, such distributions are
characterized not only by the average current and variance, but also
by higher order moments. This study of fluctuations has attracted
intense theoretical activity in the recent years, and powerful
frameworks such as full counting statistics (FCS) have been
formulated.\cite{levitov96,nazarov} Starting from fundamental
microscopic theories, moment-generating functions are known for
several systems such as tunnel junctions and quantum dots.
Experimental investigation of the fluctuations beyond the variance
is, however, not so well-established. Experiments observing higher
moments of current or voltage have remained difficult and time
consuming as filtering and bandwidth requirements are hard to
fulfill --- detection of higher moments of fluctuations is typically
characterized by very weak signals and wide bandwidth measurements
performed at low temperatures where thermal effects are suppressed.

The first measurements\cite{reulet03} of the third moment of current
fluctuations across a voltage-biased tunnel junction support the
theoretical concepts, although the use of conventional amplifiers
and mixers requires long averaging times. Furthermore, the
electrical environment of the conductor can significantly affect the
measured statistics.\cite{beenakker03} In a more recent
measurement,\cite{bomzea05} the use of a slightly different
detection scheme provides a more direct access to fluctuations of
voltage up to the third moment.

Among the growing number of experimental findings, remarkable
results have been obtained using real-time detection of single
electrons,\cite{lu03,delsing05} which can be utilized in the
measurement of full counting statistics of electron transport.
Recently, even further progress was accomplished in a measurement of
the distribution of tunneling current through a quantum
dot.\cite{gustavsson06} These innovative approaches are examples of
on-chip detection of fluctuations, which is the design goal of
future experimental schemes as well. In this way, one can perform
faster measurements on a wider bandwidth and avoid the problems with
remotely connected amplifiers.

On the other hand, the above techniques based on direct counting of
electrons are best suited up to moderate frequencies and extremely
low current levels in the pA range and below. Other experimental
approaches are needed at considerably higher currents and for
frequencies in the range of several GHz, indispensable for the
characterization of many involved processes. One solution is the use
of Josephson junctions (JJs) as noise probes. With small
low-capacitance junctions, the sensitivity of the Coulomb blockade
can be used to characterize voltage fluctuations generated by a
mesoscopic conductor.\cite{heikkila04,lindell04} Another possibility
is the use of larger current-biased junctions as threshold
detectors\cite{nazarov04,pekola04} to probe the higher moments of
current noise. In this case, switching rate of the junction from the
supercurrent to the normal state depends strongly on the
fluctuations in the bias current. The applicability of such a JJ
detector to measure shot noise has been demonstrated,\cite{pekola05}
but convincing experimental results concerning the higher moments of
fluctuations have not yet been reported.

A JJ appears to be quite an attractive threshold detector at the
first view: it detects with a certain probability a current
exceeding a given threshold, and switches consequently from the
superconducting state to a well detectable normal state. A more
detailed description of the dynamics of the junction acting as a
noise detector reveals several mechanisms though: apart from the
usual crossover from thermal activation (TA) to macroscopic quantum
tunneling (MQT) as a function of temperature, the response of a
hysteretic Josephson junction to current noise depends inherently on
both the spectrum and distribution of the fluctuations. In the TA
regime, the switching of the junction occurs by thermally activated
escape over a high potential barrier. On the contrary, at lower
temperatures the escape results as the phase of the superconducting
order parameter over the junction tunnels quantum mechanically
through the barrier. These processes are affected by current
fluctuations, and the influence of noise on the escape dynamics has
been correspondingly analyzed for junctions in different parameter
and temperature
regimes.\cite{pekola05,ankerhold05,duckheim05,ankerhold06} For
example, high-frequency noise leads to the notion of an effective
temperature $T^{*}$ of the detecting junction, which often exceeds
the superconducting transition temperature
$T_{\mathrm{c}}$.\cite{pekola05}

In this work, we first characterize in general the influence of
current fluctuations on a Josephson junction initially in the regime
of macroscopic quantum tunneling. Consequently, we discuss the
sensitivity of a JJ when detecting the higher moments of
high-frequency noise using switching measurements. This leads us to
investigate the influence of weaker low-frequency fluctuations that
can be considered adiabatic in view of the quantum dynamics of the
detector. Since achieving suitable sensitivity to fluctuations in
this certain frequency range depends strongly on the electrical
environment of the detector, we discuss the requirements for a
feasible experimental detection scheme such as increasing the plasma
frequency $\omega_{\mathrm p}$ of the detector junction or including
filtering to limit the accessible noise bandwidth to the desired
range.

\section{INFLUENCE OF CURRENT FLUCTUATIONS ON A JOSEPHSON JUNCTION DETECTOR IN DIFFERENT FREQUENCY REGIMES}

In this work we concentrate on discussing how a current-biased
hysteretic Josephson junction in the regime of macroscopic quantum
tunneling can be used to characterize weak fluctuations in its bias
current. Such a junction can act as a sensitive on-chip detector of
current noise since the tunneling rate of the phase depends
exponentially on current fluctuations. Under certain conditions this
sensitivity may allow to distinguish non-Gaussian features of the
fluctuations, which forms a central part of the discussion to
follow.

In a general situation when the detector junction is part of an
electrical circuit,
these fluctuations of bias current may be of several origins. First,
equilibrium noise is always present in the circuit, even when
external current and voltage sources are turned off. For a general
linear electrical circuit, the spectral density
$S_I(\omega)\equiv\int_{-\infty}^{\infty}\mathrm{d}t\langle
I(t)I(0)\rangle\exp(i\omega t)$ of these normally distributed
equilibrium fluctuations is determined by the
fluctuation-dissipation theorem. As in the previous formula,
assuming the distribution of the fluctuations to be stationary, the
noise power at the detector junction at frequency $\omega$ is given
by\cite{nazarov}
\begin{equation}
S_I^{\mathrm{env}}(\omega)=2\hbar\omega\mathrm{Re}[Y(\omega)]\left(\coth\frac{\hbar\omega}{2k_{\mathrm{B}}T}+1\right),
\label{s_env}
\end{equation}
where $T$ is the temperature and $Y(\omega)$ is the
frequency-dependent admittance of the electrical circuit seen from
the detector. At low frequencies or high temperatures with
$\hbar\omega\ll k_{\mathrm{B}}T$, this reduces to the familiar
expression for white thermal noise,
$S_I^{\mathrm{env}}(\omega)=4Gk_{\mathrm{B}}T$ with
$G\equiv\mathrm{Re}[Y(\omega\approx 0)]$ denoting an effective
conductance. Likewise, at the opposite limit we recover the
expression for high-frequency quantum noise.

Besides these equilibrium fluctuations, generally non-Gaussian
nonequilibrium noise may be present in the circuit as well. For
example, in a circuit containing a voltage-biased tunnel junction in
the normal state, nonequilibrium shot noise arises from the
stochastic tunneling of discrete charges through the junction.

When both equilibrium and nonequilibrium fluctuations affect the
detector, the total noise power at the detector junction is obtained
as the incoherent sum of the different contributions:
$S_I(\omega)=S_I^{\mathrm{env}}(\omega)+S_I^{\mathrm{shot}}(\omega)$.
The noise power related to the second cumulant of the shot noise at
a finite temperature in the low-frequency limit is given
by\cite{blanter}
$S_{I}^{\mathrm{shot}}=2eF_2\bar{I}_{\mathrm{N}}\coth(eV/2k_{\mathrm{B}}T)$,
where $\bar{I}_{\mathrm{N}}=V/R_{\mathrm{N}}$ is the average current
through the junction with resistance $R_{\mathrm{N}}$ at the bias
voltage $V$. Furthermore, $F_2$ is the Fano factor of the second
moment, relating the microscopic transport properties of the
conductor to the measurable noise. For a normal tunnel junction with
low barrier transparency, $F_2=1$. Similarly, the third cumulant of
these nonequilibrium fluctuations has in the same limit the non-zero
theoretical value $C_3=F_3e^2\bar{I}_{\mathrm{N}}$, where $F_3$ is
the Fano-factor of the third
moment.\cite{levitov96,levitov04,salo06} The third cumulant
describes the first-order deviations or asymmetry as compared to a
Gaussian distribution, being therefore one of the first quantities
to measure when investigating the noise properties of any mesoscopic
conductor beyond the variance of current.

\subsection{Dynamics of a Josephson junction}
The above expressions or their extensions give the spectral
densities of fluctuations at an arbitrary frequency and temperature.
The response of the detecting junction to current noise at different
frequencies can be explained by describing its dynamics using the
common resistively and capacitively shunted junction (RCSJ)
--model.\cite{stewart,mccumber,likharev} To be more specific, let us
consider a JJ with critical current $I_{\mathrm{c}}$ and capacitance
$C$ under the influence of an external bias current $I(t)=I_0+\delta
I(t)$. Here $I_0$ is the time-independent average bias current and
$\delta I(t)$ describes the fluctuations. Classically, for $I\leq
I_{\mathrm{c}}$ the JJ stays in the supercurrent state and the
voltage over the junction is zero. On the other hand, a bias current
$I>I_{\mathrm{c}}$ causes it to switch to the resistive state,
leading to a finite voltage of at least twice the superconducting
energy gap $\Delta$ to develop over the junction.

In the RCSJ-model, such a JJ is described in terms of a parallel
combination of a capacitor with capacitance $C$, an ideal tunnel
junction with critical current $I_{\mathrm{c}}$ and by a shunt
resistance $R$. The current--phase relation of the tunnel junction
follows the Josephson relation $I=I_{\mathrm{c}}\sin\varphi$, where
$\varphi$ is the phase difference of the superconducting order
parameter over the junction, related to the voltage by
$V=\hbar\dot{\varphi}/2e$.

\begin{figure}[!htb]
\centerline{\includegraphics[width=70mm]{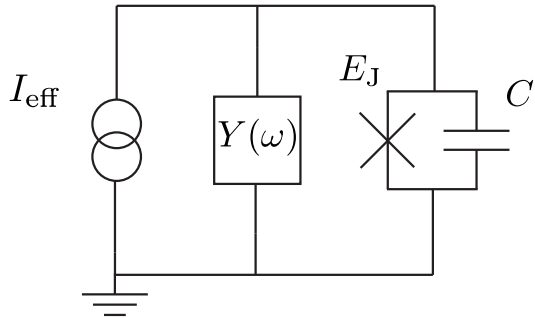}} \caption{An
effective circuit model for a JJ used as a noise detector.
Nonequilibrium fluctuations are included in the properties of the
effective current source producing a current $I_{\mathrm{eff}}$.
Electrical environment of the junction is described by the
admittance $Y(\omega)$ responsible for equilibrium noise.}
\label{fig:effective}
\end{figure}

For most of the discussion to follow, the electrical environment of
the junction can be described by the effective circuit illustrated
in Fig. \ref{fig:effective}. Here the current source
$I_{\mathrm{eff}}$ contains both a constant bias component $I_0$ and
a time-dependent part related to $\delta I(t)$, whose statistical
properties are determined by the shot noise source.

For small low-capacitance junctions the charging energy
$E_{\mathrm{C}}\equiv e^2/2C$ is not negligible, leading to a
quantum mechanical description of the junction\cite{ingold} as the
charge $q$ and phase $\varphi$ over the junction are conjugate
variables satisfying $\left[\varphi,q\right]=2ie$. In the limit of
low dissipation, the junction is described by the RCSJ-Hamiltonian
\begin{equation}
\mathcal{H}=\frac{q^2}{2C}+U(\varphi)=\frac{q^2}{2C}-E_{\mathrm{J}}\left[\cos\varphi+\frac{I(t)}{I_{\mathrm{c}}}\varphi\right],\label{hamiltonian1}
\end{equation}
where $E_{\mathrm{J}}\equiv\hbar I_{\mathrm{c}}/2e$ is the Josephson
coupling energy. With the above commutation relation in mind, the
behavior of a junction with $E_{\mathrm{C}}\ll E_{\mathrm{J}}$ can
therefore be visualized as that of an imaginary quantum particle
with mass $m=(\hbar/2e)^2C$ moving in a tilted cosine potential
$U(\varphi)$. At low temperatures, the $\varphi$-particle is
localized in one of the wells of this 'washboard' potential,
performing oscillations at a local minimum at the plasma frequency
$\omega_{\mathrm{p}}=\sqrt{8E_{\mathrm{J}}E_{\mathrm{C}}\varphi_0}/\hbar$.
Here the parameter $\varphi_0$ is related to the bias current $I$
changing the tilt of the potential by
$\varphi_0\equiv\sqrt{1-(I/I_{\mathrm{c}})^2}$.

Localization of the phase particle corresponds to the junction being
in the supercurrent state. The state is, however, metastable, and
the phase can escape from the potential well by macroscopic quantum
tunneling through the barrier. To estimate the rate of this
tunneling process, for currents $I\lesssim I_{\mathrm{c}}$ one
well-barrier section of the potential $U(\varphi)$ can be
approximated by a cubic parabola\cite{weiss}
\begin{equation}
U(\varphi)\simeq\frac{3}{4}\Delta
U\left(\frac{\varphi}{\varphi_0}\right)^2\left(1-\frac{1}{3}\frac{\varphi}{\varphi_0}\right),\label{hamiltonian2}
\end{equation}
where $\varphi$ is now measured from a minimum of the potential and
$\Delta U$ is the height of the barrier given by
\begin{equation}
\Delta
U=\frac{2}{3}E_{\mathrm{J}}\varphi_0^{3}\simeq\frac{4\sqrt{2}}{3}E_{\mathrm{J}}\left(1-\frac{I}{I_{\mathrm{c}}}\right)^{\frac{3}{2}}.\label{deltau}
\end{equation}
Likewise, the tunneling rate $\Gamma(I)$ is obtained as
\begin{equation}
\Gamma(I)\equiv
A(I)e^{-B(I)}=12\sqrt{6\pi}\frac{\omega_{\mathrm{p}}}{2\pi}\sqrt{\frac{\Delta
U}{\hbar\omega_{\mathrm{p}}}}\exp\left(-\frac{36}{5}\frac{\Delta
U}{\hbar\omega_{\mathrm{p}}}\right). \label{gamma}
\end{equation}
This follows from the treatment of tunneling out of a metastable
cubic potential well in the semiclassical limit together with
negligibly low dissipation.\cite{weiss} Such a situation arises for
small values of the real part of the effective shunting admittance
$Y(\omega)$, corresponding to a high quality factor $Q\equiv
\omega_{\mathrm{p}}C/\mathrm{Re}[Y(\omega)]$.

Assuming that a constant current $I$ is applied for a time $\Delta
t$, the finite decay rate $\Gamma(I)$ leads to the probability
\begin{equation}
P(I)=1-\exp(-\Gamma(I)\Delta t) \label{prob1}
\end{equation}
for the phase to escape from the potential well. After tunneling,
the particle starts running down the potential hill. For a
hysteretic junction with $Q\gg 1$, this corresponds to the junction
actually switching to the finite voltage state since the particle
will not be trapped again into a local minimum until the bias
current is lowered below the retrapping current close to zero.
Hence, the quantum tunneling can be experimentally observed by
applying repeated current pulses of constant height $I_0$ and length
$\Delta t$ and recording the number of times the junction has
switched to the normal state. Different values of $I_0$ then yield
an escape probability histogram $P(I_0)$ to be compared with Eq.
\equref{prob1}.

At higher temperatures, the MQT process is no longer the dominant
way of escaping from the metastable well. Instead, the energy levels
of the well have approximately thermal populations and the particle
can escape over the potential barrier by thermal activation (TA)
characterized by the rate\cite{weiss}
\begin{equation}
\Gamma_{\mathrm{T}}\simeq\frac{\omega_{\mathrm{p}}}{2\pi}\exp\left(-\frac{\Delta
U}{k_{\mathrm{B}}T}\right).\label{ta1}
\end{equation}
Here $\Delta U$ is again the height of the barrier introduced in Eq.
\equref{deltau}.

\subsection{Effect of noise on escape characteristics}
\label{sec:noiseprinciple} 
\begin{figure}[!htb]
\centerline{\includegraphics[width=120mm]{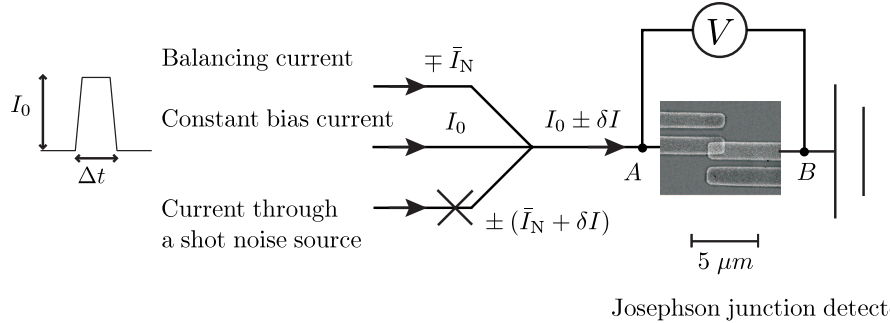}}
\caption{A general measurement scheme to detect current noise with a
JJ, showing the ideal flow of currents and a micrograph of a typical
Al--AlO$_{\mathrm{x}}$--Al JJ manufactured using electron beam
lithography and 2-angle shadow evaporation. In a typical
measurement, an average current $\bar{I}_{\mathrm{N}}$ is first
injected through the noise source (a mesoscopic scatterer), whose
temporal current can then be written as $\bar{I}_{\mathrm{N}}+\delta
I(t)$. The DC component of this current is returned through the
upper arm in the figure. In practise, this is achieved by injecting
a balancing current $-\bar{I}_{\mathrm{N}}$ through the upper line.
As a result, ideally only the fluctuating current $\delta I(t)$
flows through the detector junction located between points $A$ and
$B$ in the figure. This balance of currents is verified by
monitoring the average current through the detector. For a suitably
designed circuit, the fluctuations $\delta I(t)$ pass mainly through
the detector and do not leak back through the biasing lines.
Finally, the escape rate of the detector JJ in the presence of the
fluctuating current $\delta I(t)$ is measured by injecting $N$
pulses of constant height $I_0$ and length $\Delta t$ through a
third line (middle arm in the figure). The $I_0$--pulses have long
leading and trailing edges to ensure that the detector responds
adiabatically to them. As a result of this three-fold current
injection scheme, the detector is effectively biased by a current
$I_0+\delta I(t)$. The switching of the detector junction out of the
supercurrent state is detected as voltage pulses between points $A$
and $B$. The effects of non-zero higher odd moments of the
distribution of $\delta I(t)$ can be probed by inverting the
currents $-\bar{I}_{\mathrm{N}}$ and $\bar{I}_{\mathrm{N}}+\delta
I(t)$, as indicated by the alternative signs in the figure.
Inverting the average current through the noise source changes the
sign of the third cumulant of the fluctuations, resulting in
differing average escape rates of the detector.}\label{fig:setup}
\end{figure}

\begin{figure}[!htb]
\centerline{\includegraphics[width=120mm]{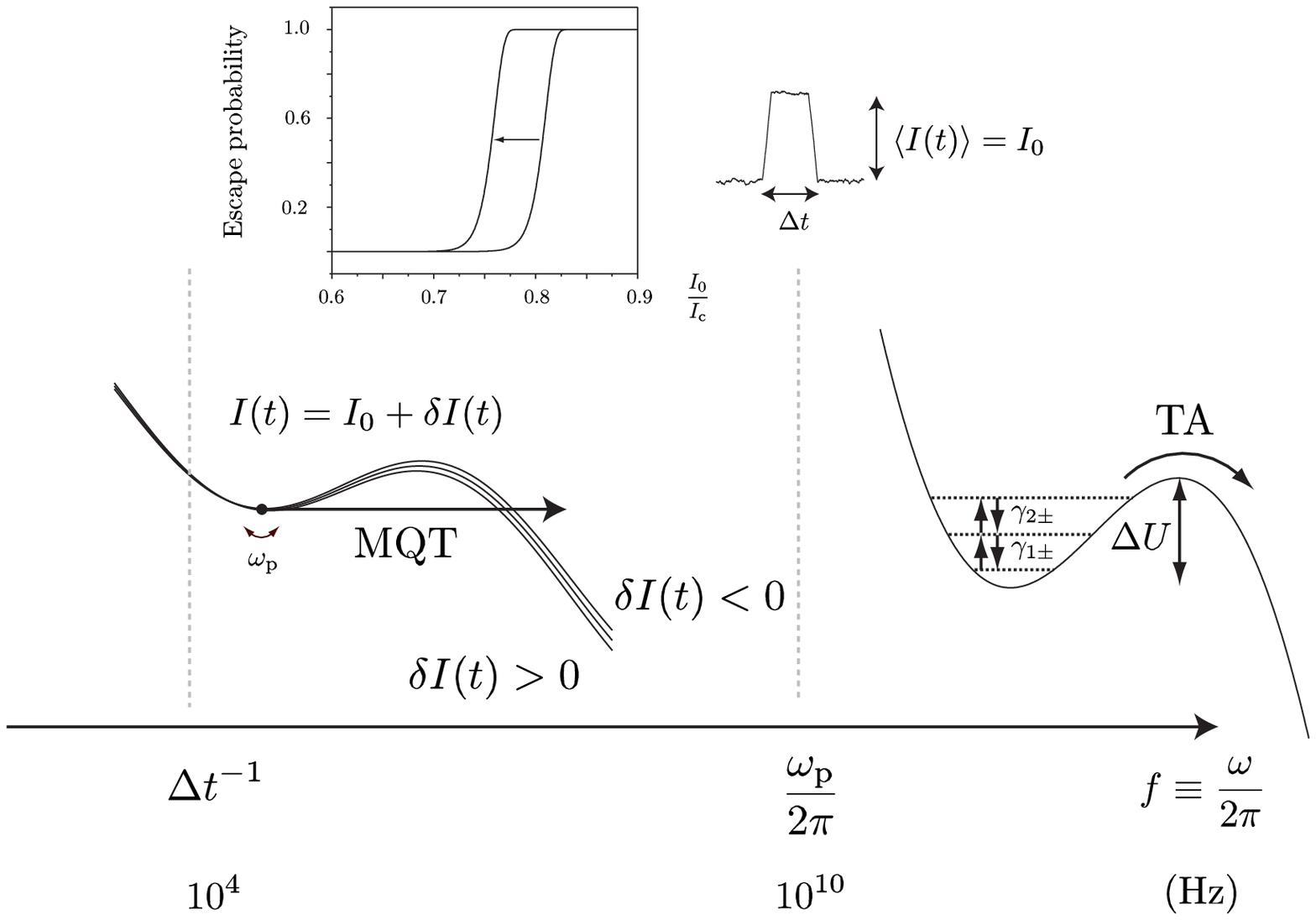}}
\caption{Typical behavior of a JJ in the presence of current
fluctuations belonging to different frequency ranges as discussed in
the text. The fluctuations $\delta I(t)$ cause the measured escape
histograms to shift toward lower values of $I_0$. In this work we
focus on the wide adiabatic regime spanning frequencies $f$
satisfying $\Delta t^{-1}\ll f\ll \frac{\omega_{\mathrm{p}}}{2\pi}$.
Here $\Delta t^{-1}$ is the inverse length of a bias current pulse
and $\omega_{\mathrm{p}}$ is the plasma frequency of the detector
junction. Current fluctuations in this regime slowly vary the tilt
of the potential, and the escape rate is obtained as an average of
the MQT rate over the distribution of the current values. On the
other hand, high frequency noise with
$f\sim\frac{\omega_{\mathrm{p}}}{2\pi}$ excites the JJ from its
ground state, leading to thermal activation at an effective
temperature.} \label{fig:freq}
\end{figure}

The above view of the dynamics of a JJ
allows us to immediately distinguish a few different frequency
regimes in terms of the response of the detecting junction to bias
current fluctuations. To see this, let us suppose that the switching
rate of the detector junction in the presence of noise is
experimentally determined using the principle presented in Fig.
\ref{fig:setup}. In more detail, we assume that $N$ constant bias
current pulses of height $I_0$ and length $\Delta t$ injected in a
time $\Delta t_{I_0}$ are used to obtain each single point on an
escape probability histogram $P(I_0)$. The number $N$ is determined
by the desired limit on the statistical error in the measurement.
Since the total bias current $I(t)=I_0+\delta I(t)$ now contains a
fluctuating part, the result is an average escape histogram
differing from the ideal curve measured without the fluctuations
$\delta I(t)$. Then, depending on the frequency content of the
fluctuating current $\delta I(t)$,
the pure MQT or TA rate [cf. Eqs. \equref{gamma} and \equref{ta1}]
is modified as illustrated by the guidelines of Fig. \ref{fig:freq},
showing the typical behavior of the junction and the relevant
characteristic frequencies.

For noise with frequencies well between $\Delta t_{I_0}^{-1}$ and
$\Delta t^{-1}$, the detector junction is effectively biased by a
constant current $I_{0,j}$ during the $j$:th pulse. Consequently,
the escape probability is approximately determined as an average
over the distribution of $I_{0,j}$ probed by the detector when
injecting $N$ current pulses to measure a single point on a
histogram. The effects of noise at these low frequencies are
considered for example in Ref. \onlinecite{buisson}, and will not be
discussed further in this work. Above this frequency range, weak
current fluctuations with frequencies up to the order of
$\omega_{\mathrm{p}}$ belong to the so-called adiabatic regime.
These quasistationary fluctuations have low enough frequencies so
that the noise $\delta I(t)$ can be considered merely as a parameter
slowly varying the tilt of the washboard potential around its
average.\cite{martinis88} In this limit the junction remains in its
instantaneous metastable quantum mechanical ground state that decays
via MQT. Since the tunneling rate is strongly dependent on current,
the junction can act as a detector of fluctuations.

Contrary to this weak low frequency noise, stronger fluctuations
with frequencies comparable to $\omega_{\mathrm{p}}$ and thus to the
level separation in the potential well excite the phase particle to
higher energy levels, leading to escape over the barrier top. Again,
the junction is sensitive to the fluctuations but this time the
noise affects the rate of effective thermal activation instead of
quantum tunneling.

Although detecting the influence of the third cumulant of adiabatic
current fluctuations is the primary goal of the measurement scheme
considered in this work, we will first focus on the opposite,
high-frequency limit.
This is necessary as the
power of noise at these high frequencies sets a condition for the
validity of the discussion in the adiabatic regime.

\section{EFFECTS OF CURRENT FLUCTUATIONS AT NON-ADIABATIC
FREQUENCIES} \label{sec:tstar}

To understand the influence of high-frequency noise on a JJ
detector, let us briefly review the model introduced in Ref.
\onlinecite{pekola05}. The effects of the second moment at high
frequencies can be modeled as resonant excitation between different
energy levels in the nearly harmonic potential well of the detector
junction. At low temperatures, this process is mainly driven by the
nonequilibrium shot noise. The transitions take place approximately
at the plasma frequency of the junction, and the resulting level
dynamics can be described in terms of an effective temperature $T^*$
leading to thermal activation over the potential barrier. In the
following, we choose to denote the excitation and relaxation rates
between the $j-1$:th and $j$:th level in the well separated by
energy $\hbar\omega_{j,j-1}$ by $\gamma_{j,j-1}$ and
$\gamma_{j-1,j}$, respectively.

To derive expressions for the transition rates, we start from the
Hamiltonian $\mathcal{H}$ of a JJ biased with a fluctuating current
$I(t)=I_0+\delta I(t)$ and write $\mathcal{H}$ in the general form
as $\mathcal{H}(t)=\mathcal{H}_0+V(t)$, where $\mathcal{H}_0$ is the
time-independent system Hamiltonian and $V(t)$ is a perturbation of
the form $g\hat{A}\hat{f}(t)$. Here $g$ is a coupling constant and
$\hat{A}$ is an operator of the system (the detector junction). The
matrix elements $\langle j|\hat{A}(t=0)|j-1\rangle$ of $\hat{A}$
between states $|j\rangle$ and $|j-1\rangle$ at $t=0$ are expressed
as $A^{j,j-1}$. Further, $\hat{f}(t)$ is a fluctuating operator
commuting with $\hat{A}$ with a power spectrum $S_f(\omega)$.

We proceed by expanding the unitary time-evolution of a state
$|\psi_I(t)\rangle$ in the interaction picture up to third order in
the perturbation $V(t)$:
\begin{align}
|\psi_I(t)\rangle\simeq\left\{1-\frac{i}{\hbar}\int_0^t\mathrm{d}\tau_1
V(\tau_1)-
\frac{1}{2\hbar^2}\int_0^t\int_0^t\mathrm{d}\tau_1\mathrm{d}\tau_2\mathcal{T}\left[V(\tau_1)V(\tau_2)\right]\right.\nonumber\\
\left.+\frac{i}{6\hbar^3}
\int_0^t\int_0^t\int_0^t\mathrm{d}\tau_1\mathrm{d}\tau_2\mathrm{d}\tau_3
\mathcal{T}\left[V(\tau_1)V(\tau_2)V(\tau_3)\right]\right\}|\psi_I(0)\rangle,\label{timedev1}
\end{align}
where the time-ordering operator $\mathcal{T}$ is needed as the
operators $V(\tau)$ at different times do not necessarily commute.
To study the influence of the second moment of fluctuations, we
neglect temporarily the last two terms on the right hand side of Eq.
\equref{timedev1}. Now, considering the time-evolution of the
initial state $|\psi_I(0)\rangle=|j\rangle$, we first find the
probability amplitude $\alpha_{j-1,j}(t)\equiv\langle
j-1|j(t)\rangle$ for a transition to state $|j-1\rangle$. The
corresponding ensemble-averaged relaxation rate follows then from
the standard expression\cite{schoelkopf}
\begin{equation}
\gamma_{j-1,j}\equiv\frac{\mathrm{d}}{\mathrm{d}t}\langle|\alpha_{j-1,j}(t)|^2\rangle=\frac{g^2}{\hbar^2}|A^{j,j-1}|^2S_f(\omega_{j,j-1}).\label{transition1}
\end{equation}
A similar calculation gives the excitation rate from the $j-1$:th to
the $j$:th level as
\begin{equation}
\gamma_{j,j-1}\equiv\frac{\mathrm{d}}{\mathrm{d}t}\langle|\alpha_{j,j-1}(t)|^2\rangle=\frac{g^2}{\hbar^2}|A^{j,j-1}|^2S_f(-\omega_{j,j-1}),\label{transition2}
\end{equation}
implying that relaxation is mainly determined by the noise power at
positive and excitation at negative frequencies.

The detector junctions usually have $E_J\gg E_C$, and we can write
the Hamiltonian $\mathcal{H}_0$ in the tight-binding approximation
of Eq. \equref{hamiltonian1} with $I=I_0$ and further approximate
the potential by a harmonic one. Referring to Eq.
\equref{hamiltonian1}, we have $g=-E_J/I_c$, $\hat{A}=\varphi$ and
$\hat{f}(t)=\delta I(t)$. The required matrix elements for the
harmonic potential are given by $j/2\hbar\omega_{\mathrm{p}}C$,
yielding
$\gamma_{j-1,j}\simeq(j/2\hbar\omega_pC)S_I(\omega_{j,j-1})$ for the
relaxation and
$\gamma_{j,j-1}\simeq(j/2\hbar\omega_pC)S_I(-\omega_{j,j-1})$ for
the excitation rate. Here, the total noise power at the detector
\begin{align}
S_I(\pm\omega_{j,j-1})&=
2eF_2\bar{I}_N\alpha\coth\frac{eV}{2k_{\mathrm{B}}T}\nonumber \\
&+2\hbar\omega_{j,j-1}\mathrm{Re}[Y(\omega_{j,j-1})]\left(\coth\frac{\hbar\omega_{j,j-1}}{2k_{\mathrm{B}}T}\pm
1\right)\label{noisepower}
\end{align}
is a sum of the contributions from the shot noise source and the
equilibrium fluctuations arising from the dissipative circuit
surrounding the junction. The former is scaled by a factor $\alpha$:
It relates the noise power at the source to the noise power at the
detector at frequency $\omega_{j,j-1}$. The frequency-dependent
factor $\alpha$ can be calculated by solving the set of Langevin
equations written for each branch of the circuit,\cite{blanter} but
for the present case we may combine it with $F_2$ into an effective
Fano factor $F$ of the complete circuit. With this in mind, the
transition rates read
\begin{align}
\gamma_{j-1,j}&\simeq
\frac{jFe\bar{I}_{\mathrm{N}}\coth(eV/2k_{\mathrm{B}}T)}{\hbar\omega_{\mathrm{p}}
C}+j\frac{\omega_{j,j-1}}{Q}\left(\coth\frac{\hbar\omega_{j,j-1}}{2k_{\mathrm{B}}T}+1\right)\nonumber\\
\gamma_{j,j-1}&\simeq
\frac{jFe\bar{I}_{\mathrm{N}}\coth(eV/2k_{\mathrm{B}}T)}{\hbar\omega_{\mathrm{p}}
C}+j\frac{\omega_{j,j-1}}{Q}\left(\coth\frac{\hbar\omega_{j,j-1}}{2k_{\mathrm{B}}T}-1\right)
.\label{relaxation1}
\end{align}
Here, $Q\equiv C\omega_{\mathrm{p}}/\mathrm{Re}[Y(\omega_{j,j-1})]$
denotes the quality factor of the detector junction. The level
dynamics following from Eq. \equref{relaxation1} is equivalent to
that arising from pure equilibrium fluctuations at an effective
temperature $T^{*}$ if we require
\begin{align}
\gamma_{j-1,j}\equiv&\frac{j\omega_{j,j-1}}{Q}\left(\coth\frac{\hbar\omega_{j,j-1}}{2k_{\mathrm{B}}T^{*}}+1\right)\nonumber\\
\mathrm{and}\quad\gamma_{j,j-1}\equiv&\frac{j\omega_{j,j-1}}{Q}\left(\coth\frac{\hbar\omega_{j,j-1}}{2k_{\mathrm{B}}T^{*}}-1\right).
\label{tstar1}
\end{align}
Comparing this expression with Eq. \equref{relaxation1}, we have,
assuming $\omega_{j,j-1}\simeq\omega_p$,
\begin{equation}
k_{\mathrm{B}}T^{*}\simeq\frac{\hbar\omega_{\mathrm{p}}}{2\mathrm{arcoth}\left(\coth\frac{\hbar\omega_{\mathrm{p}}}{2k_{\mathrm{B}}T}+
\frac{QFe\bar{I}_{\mathrm{N}}\coth(eV/2k_{\mathrm{B}}T)}{\hbar\omega_{\mathrm{p}}^2
C}\right)}. \label{tstar2}
\end{equation}
In the absence of shot noise when $\bar{I}_{\mathrm{N}}=0$, $T^{*}$
reduces to the actual temperature $T$. For bias voltages $e|V|\gg
k_{\mathrm{B}}T$ together with $\hbar\omega_{\mathrm{p}}\gg
k_{\mathrm{B}}T$, $T^{*}$ is obtained from
\begin{equation}
k_{\mathrm{B}}T^{*}\simeq\frac{\hbar\omega_p}{2\mathrm{arcoth}\left(1+\frac{QFe|\bar{I}_{\mathrm{N}}|}{\hbar\omega_p^2C}\right)}.
\label{tstar3}
\end{equation}
Consequently, for high noise currents with
$|\bar{I}_{\mathrm{N}}|\gg\hbar\omega_{\mathrm{p}}^2C/QFe$, Eq.
\equref{tstar3} gives $T^{*}\simeq
QFe|\bar{I}_{\mathrm{N}}|/2k_{\mathrm{B}}\omega_{\mathrm{p}}C=Fe|\bar{I}_{\mathrm{N}}|/2\mathrm{Re}[Y(\omega_{\mathrm{p}})]$.
On the other hand, if the noise currents are still in the high limit
but $\hbar\omega_{\mathrm{p}}\lesssim 2k_{\mathrm{B}}T$, the
effective temperature contains also a term proportional to
temperature $T$:
\begin{equation}
k_{\mathrm{B}}T^{*}\simeq
k_{\mathrm{B}}T+\frac{QFe|\bar{I}_{\mathrm{N}}|}{2\omega_{\mathrm{p}}C}.\label{tstar4}
\end{equation}
A similarly defined effective temperature is employed also in Ref.
\onlinecite{ankerhold06}.

Now, if $T^{*}$ exceeds the crossover temperature for quantum
tunneling, i.e., $T^{*}>T_0\equiv\hbar\omega_p/2\pi k_{\mathrm{B}}$,
the decay of the metastable state occurs primarily via thermal
activation over the potential barrier.\cite{weiss} Assuming a bias
current $I_0$ close to $I_{\mathrm{c}}$, the potential can again be
approximated by a cubic parabola and the escape rate $\Gamma$ is
obtained from \equref{ta1} with $T=T^{*}$. The switching probability
of the junction for a bias current pulse of height $I_0$ and length
$\Delta t$ is correspondingly given by
\begin{equation}
P(I_0)=1-\exp\left(-\Gamma(I_0)\Delta t\right),
\end{equation}
allowing comparison with experimental switching histograms at
different values of $\bar{I}_{\mathrm{N}}$. For example, the $I_0$
values corresponding to 50\% escape probability, $P(I_0)=0.5$, or
the difference in $I_0$ between the 10\% and 90\% points allow for a
straightforward comparison of the theoretical model with
experimental data.

Besides describing the influence of non-adiabatic fluctuations, the
effective temperature sets a limit for the applicability of the
adiabatic model. To find the noise current $|\bar{I}_{N,0}|$
corresponding to the crossover temperature $T_0$ we require that
\begin{equation}
1+\frac{QFe|\bar{I}_{\mathrm{N},0}|}{\hbar\omega_{\mathrm{p}}^2C}=\coth\pi,\label{crossovercurrent1}
\end{equation}
by use of which we can estimate that the currents
$|\bar{I}_{\mathrm{N}}|$ have to stay considerably below
\begin{equation}
|\bar{I}_{\mathrm{N},0}|\equiv\frac{\eta\hbar\omega_{\mathrm{p}}^2C}{QFe}=\frac{2\eta
I_{\mathrm{c}}\varphi_0}{QF},\quad\eta\equiv \coth\pi-1\sim
0.0037\label{ibar}
\end{equation}
for MQT to be the main escape mechanism. This can be used as a first
approximation of the limit of validity for an adiabatic description
of the bias current fluctuations. Imposing such a limit is necessary
since the effects of the third and higher moments are masked by the
second-moment induced high frequency effects. To keep the currents
$\bar{I}_{\mathrm{N}}$ in an experimentally reasonable regime, the
factor $QF\equiv
F_2\alpha(\omega_{\mathrm{p}})C\omega_{\mathrm{p}}/\mathrm{Re}[Y(\omega_{\mathrm{p}})]$
in Eq. \equref{ibar} can be adjusted by using a desired filtering
circuit, as will be discussed in Sec. \ref{sec:filter} As an
example, let us consider a detector with $I_{\mathrm{c}}=5$
$\mathrm{\mu A}$ and $C=40$ fF which could well correspond to the
parameters of the junction in Fig. \ref{fig:setup}. If the junction
is biased at $I_0=0.5I_{\mathrm{c}}$ and the limiting noise current
$I_{\mathrm{N},0}$ is required to be 5 $\mathrm{\mu A}$ for typical
values $Q\simeq 10$ and $F_2=1$, the factor
$\alpha(\omega_{\mathrm{p}})$ related to the filtering should be of
the order of $10^{-3}$.

\subsection{Higher order effects in the non-adiabatic frequency
regime} \label{sec:na-approach} When deriving expressions for the
above relaxation and excitation rates, we assumed a harmonic
potential. The corrections to the transition rates arising from the
anharmonicity of the cubic potential can be handled in perturbation
theory. Moreover, the influence of higher cumulants of current
fluctuations is revealed by taking into account more terms in the
expansion of Eq. \equref{timedev1}. Yet, in the previous switching
measurements\cite{pekola05} these non-Gaussian features were masked
by the effects of the second moment, well described by the above
model with an effective temperature. Other experimental detection
schemes considering JJs as detectors of high frequency fluctuations
may thus become more suitable in the non-adiabatic frequency regime
when focusing on the higher order effects.

The weakness of the third-order effects is evident if we consider
all the terms in Eq. \equref{timedev1} and in addition account for
the anharmonicity of the cubic potential in first-order perturbation
theory. Using the scattering matrix-based in-out
ordering\cite{salo06} of the current operators, the averaged
time-ordered products of the operators can be expressed in terms of
three-current correlators or their spectral functions analogous to
the noise power. Assuming these functions to be
frequency-independent for the range of relevant frequencies, we find
in the zero-temperature limit that the shot-noise induced transition
rate from the initial $j$:th state to the final $j+1$:th state is
given by
\begin{equation}
\gamma_{j+1,j}=(j+1)\frac{Fe|\bar{I}_{\mathrm{N}}|}{\hbar\omega_{\mathrm{p}}C}.\label{transition02}
\end{equation}
This is nothing but the first term in Eq. \equref{relaxation1},
arising from a second order contribution in a harmonic potential ---
the third order contribution vanishes identically. On the other
hand, an initial superposition state like
$\kappa_j|j\rangle+\kappa_{j+1}|j+1\rangle$ leads to oscillating
populations of the different energy levels: The frequency of these
oscillations corresponds to the level separation $\omega_{01}$,
whereas their amplitude is proportional to the third moment of the
noise source. For a noise source with Fano factor $F_3$, a part of
this third order contribution originates from a term $\langle
j+1|\mathcal{T}\left[V(\tau_1)V(\tau_2)V(\tau_3)\right]|j\rangle$.
This contributes to the occupation probability of the $j+1$:th level
by
\begin{align}
&\int_0^t\int_0^t\int_0^t\mathrm{d}\tau_1\;\mathrm{d}\tau_2\;\mathrm{d}\tau_3\;\langle
j+1|\mathcal{T}\left[V(\tau_1)V(\tau_2)V(\tau_3)\right]|j\rangle\nonumber
\\
&=F_3e^2\alpha_0^3\frac{(j+1)^{3/2}}{\hbar^3\omega_{01}}\bar{I}_{\mathrm{N}}\left[1-\cos\left(\omega_{01}t\right)\right],\label{transition3}
\end{align}
where
$\alpha_0\equiv\frac{\hbar}{2e}(\frac{2E_{\mathrm{C}}}{E_{\mathrm{J}}})^{1/4}$.
Correspondingly, the average transition rate arising from this term
vanishes, predicting that there should be no detectable net signal
from the third order resonant transitions in the switching
measurement scheme discussed above.

Considering the other recently proposed approaches to detecting
higher-order cumulants, Ankerhold\cite{ankerhold06} has used an
effective Fokker-Planck equation to describe the influence of weak
short-correlated non-Gaussian fluctuations on a Josephson junction
in the regime of thermal activation. The detection is again based on
the sensitivity of the switching rate out of the zero-voltage state
to variations in current, and the principle of Fig. \ref{fig:setup}
can be used to reveal the presence of higher odd cumulants.

Another approach has been to consider a noise source capacitively
coupled to a small JJ in the regime of incoherent Cooper pair
tunneling.\cite{heikkila04} The third cumulant of high frequency
current fluctuations should then be detectable by comparing the
$I$--$V$ -curves of the detector junction at different values of the
noise current. On the other hand, in Ref. \onlinecite{brosco06} the
authors investigate the influence of third moment of current
fluctuations on a two level system. This detector can be
experimentally realized as a hysteretic JJ SQUID. The master
equation approach considers the time-evolution of the reduced
density matrix of the two level system and predicts third-moment
induced coherent oscillations between different states, observable
for example by studying Rabi oscillations of the system. This result
is consistent with Eq. \equref{transition3}, bearing in mind that
Eq. \equref{transition3} is obtained for a multilevel system, not a
qubit. A similar master equation approach was employed also in Ref.
\onlinecite{ojanen06} to calculate transition rates induced by the
third cumulant.

Furthermore, a measurement with a mesoscopic conductor parallel to a
current-biased JJ has been predicted to reveal the fourth cumulant
of current fluctuations.\cite{ankerhold05} This is based on a
modification of the rate of macroscopic quantum tunneling by the
mesoscopic conductor. The principle is somewhat similar to detecting
the influence of the third cumulant in the regime of adiabatic
fluctuations, as we will discuss below in more detail.

\section{INFLUENCE OF THE THIRD CUMULANT IN THE ADIABATIC
FREQUENCY REGIME} \label{sec:adiabatic}

The above model with the effective temperature relied heavily on
excitation and relaxation between different energy levels in the
potential well of the detector junction. In this section we
concentrate on fluctuations of lower frequencies that do not excite
the detector to higher levels.

To analyze the adiabatic fluctuations quantitatively, let us first
recall that the current distribution $\rho(\delta I)$ around the
average current $I_0$ is obtained as the Fourier transform
$\rho(\delta I)=\frac{1}{2\pi}\int_{-\infty}^{\infty}
\mathrm{d}ke^{-ik\delta I}\phi_{\delta I}(k)$ of the characteristic
function $\phi_{\delta I}(k)$. With $c_n$ denoting the cumulants of
the current, the characteristic function can be expressed as
$\phi_{\delta I}(k)=\exp(\sum_{n=2}^\infty \frac{(ik)^n}{n!}c_n)$.
Since the fluctuations are centered around the average current,
$\langle\delta I \rangle=c_1=0$. Consequently, the second and third
cumulants are given by $c_2=\langle \delta I(t)^2 \rangle$ and
$c_3=\langle \delta I(t)^3 \rangle$, and for simplicity we assume
the distribution to be stationary. Further, assuming the fourth and
higher cumulants to be negligibly small, we concentrate on the
influence of the third cumulant. Truncating the sum in the
characteristic function at $n=3$, one finds the corresponding
approximation to the probability density as
\begin{equation} \label{rho}
\rho(\delta I)\simeq \frac{1}{\sqrt{2\pi
c_2}}\left(1-\frac{c_3}{2c_2^2}\delta I+ \frac{c_3}{6c_2^3}\delta
I^3\right)\exp\left(-\frac{\delta I^2}{2c_2}\right).
\end{equation}
This result is valid when the skewness $\gamma$ of the current
distribution is small, i.e., when $\gamma \equiv c_3/c_2^{3/2} \ll
1$.\cite{cumulantnote}

Now, for a JJ biased by a current $I$ close to the critical current
$I_{\mathrm{c}}$, the MQT rate $\Gamma(I)$ is given by Eq.
\equref{gamma}. Writing the current as $I(t)=I_0+\delta I(t)$, the
escape probability for a current pulse of height $I_0$ and length
$\Delta t$ is obtained from
\begin{equation}
P(I_0)=1-\exp\left(-\int_{0}^{\Delta t}\Gamma(I_0+\delta
I(t))\mathrm{d}t\right). \label{prob}
\end{equation}
For adiabatic current fluctuations $\delta I(t)$ with frequencies
above the inverse pulse length $\Delta t^{-1}$ but well below the
plasma frequency $\omega_{\mathrm{p}}/2\pi$, we can assume the
fluctuations to be ergodic. In this case the time average of the
escape rate can be replaced by an ensemble average over the
distribution of fluctuations, giving $\langle \Gamma \rangle$ as
\begin{equation} \label{avedef}
\langle \Gamma \rangle \simeq \int_{-\infty}^{\infty}d\delta I
\rho(\delta I)\Gamma(I_0+\delta I).
\end{equation}
This corresponds further to an escape probability
\begin{equation}
P(I_0)=1-\exp(-\langle\Gamma\rangle\Delta t), \label{pesc}
\end{equation}
which can be directly compared with experimental escape histograms.

\begin{figure}[!htb]
\centerline{\includegraphics[width=90mm]{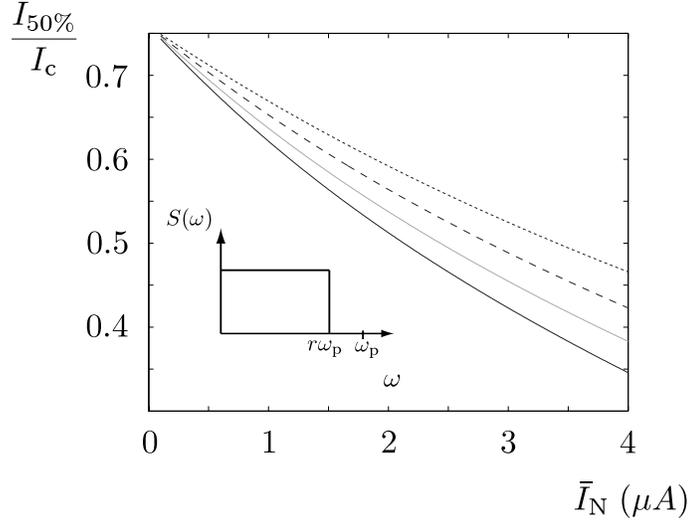}} \caption{The
bias current $I_0$ corresponding to a 50\% switching probability as
a function of $\bar{I}_{\mathrm{N}}$, the average current through
the noise source with a third cumulant $c_3=0$ within the adiabatic
model. The detector is assumed to have $I_{\mathrm{c}}=2$
$\mathrm{\mu}$A and $C=20$ fF. Furthermore, the circuit around the
JJ is assumed to have a flat frequency response up to a cutoff at
$r\omega_{\mathrm{p}}/2\pi$, as shown in the inset. The curves from
top to bottom correspond to different values of $r$ between 0.5 and
0.8 in steps of 0.1.} \label{fig:i50_in}
\end{figure}

For data fitting we obtain values of $\langle \Gamma \rangle$ by
numerical integration but for illustration we can write a second
order approximation in $\delta I$ for $\Gamma(I_0+\delta
I)=\exp\left[\ln A(I_0+\delta I)-B(I_0+\delta I)\right]$, giving
together with Eqs. \equref{rho} and \equref{avedef} the result
\begin{equation}
\langle\Gamma\rangle\simeq\frac{\Gamma_0}{\sqrt{\hat{B}c_2}}\exp\left(\frac{\hat{A}^2}{2\hat{B}}\right)
\left[1-\frac{c_3}{2c_2^2}\frac{\hat{A}}{\hat{B}}+\frac{c_3}{6c_2^3}\left(\frac{\hat{A}}{\hat{B}}\right)^3+\frac{c_3}{2c_2^3}\frac{\hat{A}}{\hat{B}^2}\right].
\label{gamma_ave2}
\end{equation}
Here $\Gamma_0$ is the unperturbed tunneling rate given by Eq.
\equref{gamma} with $I=I_0$ and we introduced the
parameters
\begin{equation} \hat{A}\equiv\frac{\partial}{\partial
I}\bigg[\ln A(I)-B(I)\bigg]_{I=I_0} \label{gamma_ave3}
\end{equation}
and
\begin{equation}
\hat{B}\equiv\frac{1}{c_2}-\frac{\partial^2}{\partial I^2}\bigg[\ln
A(I)-B(I)\bigg]_{I=I_0}. \label{gamma_ave4}
\end{equation}
This approximation is naturally valid only for small variations of
current. Moreover, neglecting the current dependence of the
prefactor $A$ and taking only first order corrections in the
exponent $B(I)$ into account, Eq. \equref{gamma_ave2} reduces to
\begin{equation}
\langle\Gamma\rangle\simeq\Gamma_0\exp\left[\frac{1}{2}\left(\frac{\partial
B}{\partial I}\right)^2 \langle \delta I^2\rangle\right]
\left[1-\frac{1}{6}\left(\frac{\partial B}{\partial I}\right)^3
\langle\delta I^3\rangle\right].\label{gamma_ave5}
\end{equation}
\begin{figure}[!htb]
\centerline{\includegraphics[width=90mm]{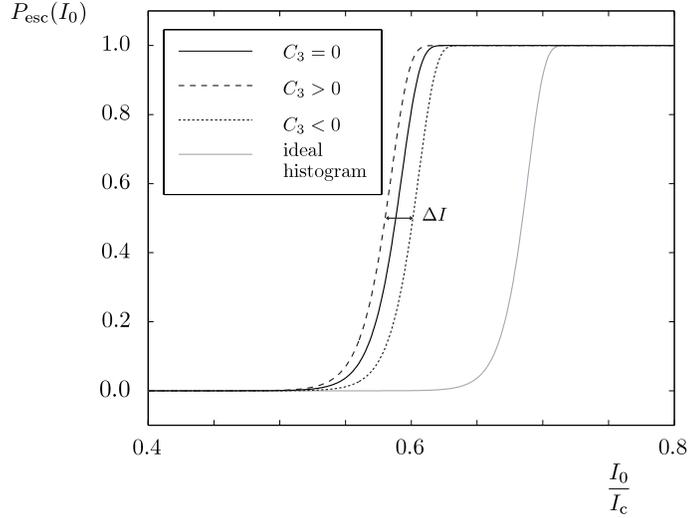}}
\caption{An example of the shift of the escape histograms compared
to the ideal noiseless case when adiabatic fluctuations are present
at the detector junction. The left group of three curves corresponds
to a Gaussian noise source with $c_3=0$ and a non-Gaussian source
with equal $c_2$ but $c_3\neq0$. Additionally, as discussed below in
more detail, the arrow indicates the shift $\Delta I$ in bias
currents corresponding to a fixed threshold probability (50\%) for
opposite signs of $c_3$. The detector is assumed to have
$I_{\mathrm{c}}=1$ $\mathrm{\mu A}$ and $C=20$ fF, whereas $c_2$ and
$c_3$ are calculated for a current $\bar{I}_{\mathrm{N}}=0.5$
$\mathrm{\mu A}$ in an ideal circuit with a cutoff frequency at 0.5
$\omega_{\mathrm{p}}$.}\label{fig:histograms}
\end{figure}
When the approximation of Eq. \equref{rho} is valid, we see that the
main effect of a nonzero current $\bar{I}_N$ through the shot noise
source is to change the average tunneling rate, resulting in a shift
of the escape probability histogram. This is illustrated in Fig.
\ref{fig:histograms} for a Gaussian source and two non-Gaussian
sources with third cumulants of opposite signs. The current
corresponding to a fixed switching probability is clearly reduced
compared to the noiseless histogram. In Fig. \ref{fig:i50_in} such
current suppression is shown for the current corresponding to 50\%
switching probability when $\bar{I}_{\mathrm{N}}$ is varied.

For different polarities of $\bar{I}_N$, the change in
$\langle\Gamma\rangle$ has the same magnitude but different
direction. In first approximation, this effect is caused solely by
the third cumulant. For numerical evaluation of the escape rate
asymmetry between different polarities of $\bar{I}_N$, we obtain
\begin{align} \label{asymmg}
\langle \Gamma\rangle_\pm &=
\frac{1}{\sqrt{2\pi}}\int_{-\infty}^{\infty} dx \Gamma_{\mathrm
0}(I_0+\sqrt{c_2}x)\exp(-x^2/2)\nonumber \\
&\pm \frac{\gamma}{2} \frac{1}{\sqrt{2\pi}}\int_{-\infty}^{\infty}
dx \Gamma_{\mathrm 0}(I_0+\sqrt{c_2}x)x(1-x^2/3)\exp(-x^2/2),
\end{align}
which follows from Eqs. \equref{rho} and \equref{avedef}. This can
then be used to evaluate the relative asymmetry of
$\langle\Gamma\rangle$ between different polarities of $\bar{I}_N$,
defined by the expression
\begin{equation} \label{deltagdef}
\frac{\Delta \Gamma}{\Gamma_{\mathrm{ave}}}\equiv \frac{\langle
\Gamma\rangle_+-\langle \Gamma\rangle_-}{\frac{1}{2}(\langle
\Gamma\rangle_++\langle \Gamma\rangle_-)}.
\end{equation}
\begin{figure}[!htb]
\centerline{\includegraphics[width=90mm]{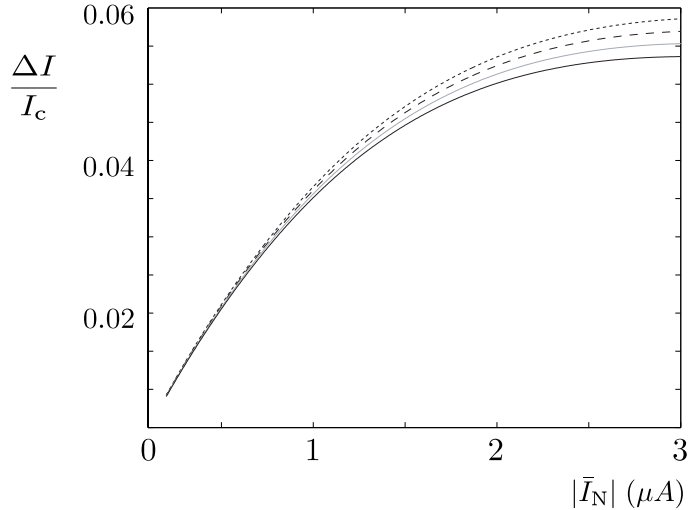}} \caption{The
shift $\Delta I/I_{\mathrm{c}}$ of a threshold current $I_{x}$ as a
function of $|\bar{I}_{\mathrm{N}}|$, the magnitude of the average
current through the noise source. $\Delta I$ is the difference of
the bias currents $I_0$ corresponding to an escape probability $P=x$
when the polarity of $\bar{I}_{\mathrm{N}}$ is changed. The
different curves from top to bottom correspond to values of $x$
between 0.5 and 0.8 in steps of 0.1. The detector is again assumed
to have $I_{\mathrm{c}}=1$ $\mathrm{\mu}$A and $C=20$ fF. In
addition, the circuit around the JJ has a flat response up to a
cutoff at $r\omega_{\mathrm{p}}/2\pi$ with $r=0.5$.}
\label{fig:deltai}
\end{figure}

The above result can further be used when approximating how a
certain point on the histogram corresponding to a fixed switching
probability moves as the sign of $\bar{I}_{\mathrm{N}}$ is changed.
This shift $\Delta I$ of a threshold current $I_0$ is illustrated in
Fig. \ref{fig:deltai}. The value of $\Delta I$ is numerically easily
obtained, e.g., by using a bisection method. In the linear
approximation, this shift due to the third moment is determined by
the combined effect of the asymmetry described by Eq.
\equref{deltagdef} and the slope of the histogram at the average
current $I_0$, $\partial P/\partial I|_{I=I_0}$. Thus, for $\Delta
I\ll$ the width of the histogram, we find
\begin{equation} \label{DeltaI}
\Delta I \simeq (1-P)\ln(1-P)\left(\frac{\partial P}{\partial
I}\right)^{-1} \frac{\Delta \Gamma}{\Gamma_{\mathrm{ave}}}
\end{equation}
for the shift in the threshold current $I_0$ as $\bar{I}_N$ is
changed to $-\bar{I}_N$.

\section{EXPERIMENTAL REALIZATION}
\begin{figure}[!htb]
\centerline{\includegraphics[width=100mm]{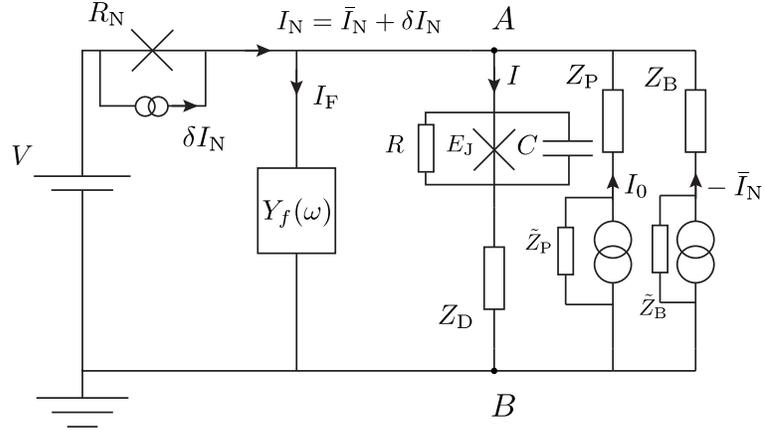}}
\caption{Circuit model for the electrical environment of the JJ used
as a noise detector. The impedances $Z_{\mathrm{D}}$,
$Z_{\mathrm{P}}$ and $Z_{\mathrm{B}}$ are mainly inductive whereas
$\tilde{Z}_{\mathrm{P}}$ and $\tilde{Z}_{\mathrm{B}}$ are
capacitive. The bias currents $I_0$ and $-\bar{I}_{\mathrm{N}}$ are
realized by applying a bias voltage over a large resistor. When the
detecting junction switches to the normal state, voltage pulses are
measured, e.g., between points $A$ and $B$.} \label{fig:filtersetup}
\end{figure}

In Sec. \ref{sec:noiseprinciple} and in Fig. \ref{fig:setup} we
briefly described the principle of detecting current fluctuations
using a Josephson junction. In this section we aim to give a more
detailed view, focusing on the electrical circuit model illustrated
in Fig. \ref{fig:filtersetup}.

The circuit is generally fabricated on a silicon substrate using
standard electron beam lithography and shadow evaporation of
aluminium. For measurements the circuit is cooled in a dilution
refrigerator down to low temperatures ($T<50$mK), where MQT is the
dominating escape mechanism in the detector, as desired. The shot
noise source is typically another small superconducting tunnel
junction in the normal state biased by a constant voltage $V$. In
practise, the noise source is typically realized as another
Josephson junction with a critical current much smaller than that of
the detector junction. Referring to Fig. \ref{fig:setup}, the
average current through this mesoscopic scatterer is given by
$\bar{I}_{\mathrm{N}}=V/R_{\mathrm{N}}$, where $R_{\mathrm{N}}$ is
the resistance of the junction at the bias voltage $V$. For
convenience, we separate the DC and AC parts of the current
$I_{\mathrm{N}}$ as $I_{\mathrm{N}}=\bar{I}_{\mathrm{N}}+\delta
I_{\mathrm{N}}$. Without extra considerations, a part of or all of
the DC current $\bar{I}_{\mathrm{N}}$ will flow through the
detector. Therefore, in a typical measurement this DC component of
the current flowing through the scatterer junction is balanced by
applying a constant current $-\bar{I}_{\mathrm{N}}$ through an
inductive line. This current bias is realized by applying a voltage
over a large resistance at room temperature ($R\geq$ 1 M$\Omega$).
As a result, no DC current flows through the detector junction until
a current pulse $I_0$ is injected through a similarly prepared line
as for the balancing current $-\bar{I}_{\mathrm{N}}$. The above
procedure allows a more accurate detection of the fluctuations for
different directions of $\bar{I}_{\mathrm{N}}$, and the balance of
DC currents is adjusted and monitored for each value of $V$ (or
$\bar{I}_{\mathrm{N}}$).

After the circuit has been balanced, a current pulse $I_0$ is
applied. The current $I_0$ has the form of a trapezoidal pulse with
height $I_0$. The pulse has long leading and trailing edges to
ensure that the detector responds adiabatically to the changing bias
current. Typical pulse lengths vary between 100 $\mu$s and 10 ms,
whereby applying a pulse corresponds to biasing the detector with a
constant current $I_0$. Most importantly, the fluctuations caused by
the shot noise source and possibly attenuated by a low-pass filter
pass primarily through the detector because of the inductive
filtering ($Z_{\mathrm{P}}$ and $Z_{\mathrm{B}}$) of the lines with
the current sources. These fluctuations imposed on top of the
constant bias $I_0$ can then be probed by sending a large number $N$
(typically $>1000$) of bias pulses of the above kind at a fixed
value of $I_0$ repeatedly through the detector to produce escape
histograms.

The filter circuit with admittance $Y_f(\omega)$ in Fig.
\ref{fig:filtersetup} is essential for guaranteeing the adiabaticity
of the current fluctuations, as we will discuss in the next section.
However, it does not qualitatively affect the measurement process
described above. On the other hand, the effective admittance
$Y(\omega)$ parallel to the detector junction gives an additional
contribution to the tunneling exponent and prefactor of Eq.
\equref{gamma}. To first order, the change of the tunneling exponent
$B$ can be taken into account for arbitrary $Y(\omega)$ following
Refs. \onlinecite{leggett} and \onlinecite{esteve}. This results in
a correction term $\Delta B$, which is, however, independent of the
bias current through the junction. Moreover, the electrical circuit
around the detector affects the cumulants $c_2$ and $c_3$ appearing
in Eq. \equref{rho}. These are related to the bias current
fluctuations over the detector $\Delta I_{\mathrm{D}}$ and their
distribution, which generally differs from that of the fluctuations
$\delta I_{\mathrm{N}}$ at the noise source.
\begin{figure}[!htb]
\centerline{\includegraphics[width=100mm]{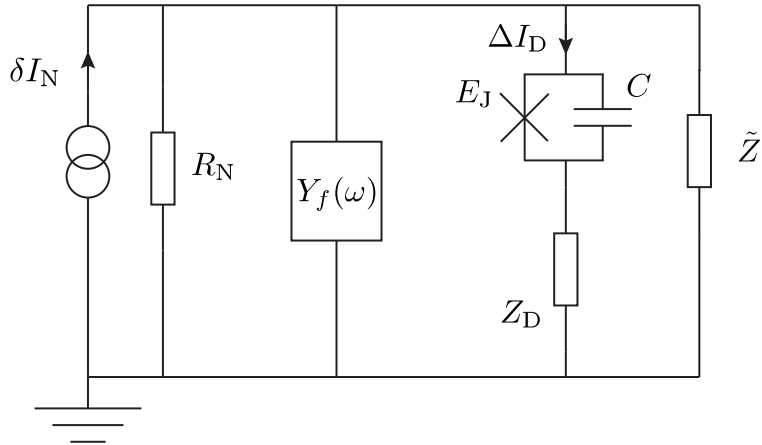}}
\caption{The circuit of Fig. \ref{fig:filtersetup} at AC
frequencies. The impedance $\tilde{Z}$ represents the biasing and
balancing circuit with the DC sources removed. The fluctuating
current $\delta I_{\mathrm{N}}$ can be related to $\Delta
I_{\mathrm{D}}$ by solving the set of Langevin equations as
explained in the text.} \label{fig:noise}
\end{figure}

At AC currents, the circuit appears as in Fig. \ref{fig:noise}.
Modeling the noise source as a current generator with current
$\delta I_{\mathrm{N}}$ parallel with a resistance $R_{\mathrm{N}}$,
we have to relate the varying current $\Delta I_{\mathrm{D}}$
through the detector back to $\delta I_{\mathrm{N}}$. After this,
the minimal-correlation estimates for the cumulants can be
calculated by multiplication and subsequent integration over the
relevant frequency interval. To achieve this, the current $\Delta
I_{\mathrm{D}}$ is expressed in terms of $\delta I_{\mathrm{N}}$ by
solving the set of Langevin equations of the circuit\cite{blanter}:
For each branch $j$, we first write the total current fluctuation as
$\Delta I_j(\omega)=\Delta V_j(\omega)Y_j(\omega)+\delta
I_j(\omega)$. Here $Y_j(\omega)$ is the admittance of the single
circuit element in branch $j$. This separates the contributions of a
fluctuating potential drop $\Delta V_j(\omega)$ and those of an
actual noise current source, $\delta I_j(\omega)$. For the circuit
of Fig. \ref{fig:noise}, the only non-zero $\delta I_j(\omega)$-term
corresponds naturally to $\delta I_{\mathrm{N}}$. The set of these
Langevin equations can then be written as a matrix equation, whose
solution gives for example the desired relation between $\Delta
I_{\mathrm{D}}$ and $\delta I_{\mathrm{N}}$. This procedure
corresponds to calculating the factor $\alpha$ appearing in the
effective temperature model discussed in Sec. \ref{sec:tstar}
Furthermore, the relations thus obtained can be used to study
additional corrections to the third moment arising from the
electrical circuit and the second moment of
fluctuations.\cite{nagaev02,beenakker03}

\section{FILTERING REQUIREMENTS FOR THE MEASUREMENT OF ADIABATIC
FLUCTUATIONS} \label{sec:filter} Above we have analyzed the effects
on the probability histograms arising from the third cumulant of
current fluctuations. Adiabaticity of the fluctuations is an
essential requirement for the theoretical model of Sec.
\ref{sec:adiabatic} to be applicable at all. Low temperature ensures
thermal excitation to be negligible so that the detector junction
stays in its metastable ground state in the absence of
nonequilibrium fluctuations. On the other hand, in the presence of a
shot noise source we have to ensure that the electrical circuit
surrounding the detector junction behaves in a desirable way at high
frequencies. If the noise power is considerable at frequencies near
$\omega_{\mathrm{p}}$ we cannot apply the adiabatic model.

\begin{figure}[!htb]
\centerline{\includegraphics[width=70mm]{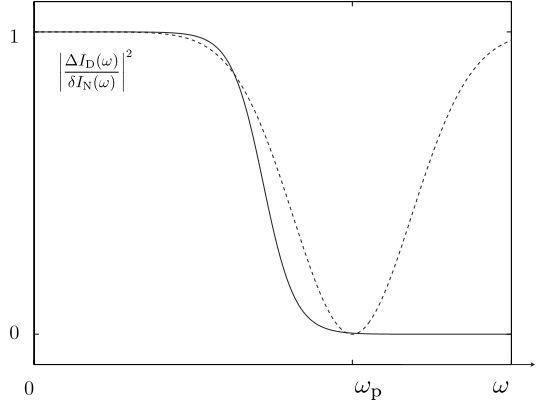}}
\caption{Idealized frequency responses of the on-chip filtering
circuit in case of a low-pass filter (solid) or a band-stop filter
(dashed). The vertical scale corresponds to the squared ratio of the
current fluctuation $\Delta I_{\mathrm{D}}(\omega)$ at the detector
junction to the current fluctuation $\delta I_{\mathrm{N}}(\omega)$
at the noise source, as illustrated in Fig. \ref{fig:noise}. The
characteristics are adjusted so that fluctuations at the plasma
frequency are suppressed by a desired amount.} \label{fig:filter}
\end{figure}

To use the adiabatic model over a wider range of
$\bar{I}_{\mathrm{N}}$, a filtering circuit has to be engineered to
suppress the high frequency nonequilibrium fluctuations. This is
illustrated in Fig. \ref{fig:filtersetup} by the filter described
with a frequency-dependent admittance $Y_f(\omega)$. Ideally, the
frequency response of the filter and the rest of the circuit
surrounding the detector, i.e., the total effective admittance
$Y(\omega)$, should resemble that of a low-pass filter with a sharp
cut-off below $\omega_{\mathrm{p}}$. Due to the admittance
$Y(\omega)$, the spectral density of fluctuations at the detector
should behave ideally as in Fig. \ref{fig:filter}. Since
$\omega_{\mathrm{p}}$ lies generally in the range of 10-100 GHz and
the filter circuit has to cover effectively a large bandwidth, the
high frequencies require us to model the circuit using techniques of
microwave engineering instead of relying merely on a lumped element
analysis.

In practise, a band reject filter with a notch around
$\omega_{\mathrm{p}}$ is easier to implement than a good low-pass
filter. This frequency response will suppress partially the
transitions caused by the second moment provided the rejection band
$\Delta\omega$ is wider than the plasma-resonance,
$\Delta\omega>\omega_{\mathrm{p}}/Q$. Moreover, if the detector
junction is replaced by a low-inductance DC SQUID with a tunable
plasma frequency, performing noise spectroscopy of the shot noise
source or characterization of the filter circuit becomes possible.
\begin{figure}[!htb]
\centerline{\includegraphics[width=140mm]{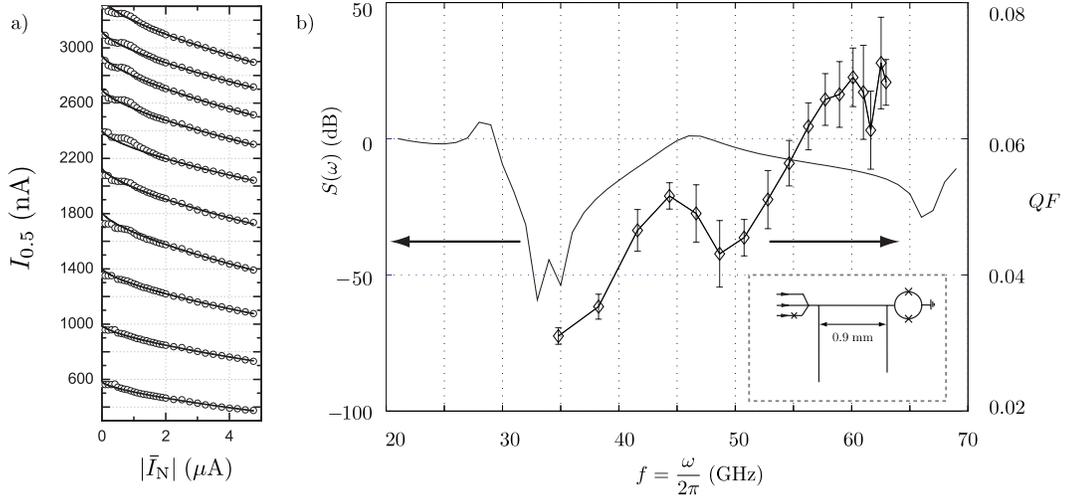}}
\caption{$\left.\mathrm{a}\right)$ Measured $I_{0.5}$ values (open
circles) as a function of $|\bar{I}_{\mathrm{N}}|$ together with
theoretical curves calculated using the effective temperature
model\cite{pekola05} with $QF$ as a fitting parameter. Different
curves correspond to different values of magnetic flux through the
SQUID loop. The experimental points below 1 $\mathrm{\mu}$A lie
almost at a constant level because the noise source is in the
superconducting state. $\left.\mathrm{b}\right)$ The $QF$--values
obtained from the fits in $\left.\mathrm{a}\right)$ as a function of
$\omega_{\mathrm{p}}$ (right scale). The factor $QF$ is partially
determined by the frequency response of the on-chip filtering
circuit, the theoretical behavior of which is shown as the solid
line (left scale). The inset shows a schematic of the sample (cf.
Fig. \ref{fig:setup}), in which the junction acting as a noise
source and the detector SQUID are separated by a band-stop filter.}
\label{fig:response}
\end{figure}

Considering the experimental realization of such a filter, we
manufactured a sample containing a JJ acting as the noise source and
a JJ detector separated from the source by a band-stop filter with a
stopband centered around 30-40 GHz. The theoretical frequency
response of the circuit is shown as the solid line in Fig.
\ref{fig:response} b), whereas Fig. \ref{fig:response} a) shows
experimental results of the current $I_{0.5}$ corresponding to 50\%
switching probability as a function of $|\bar{I}_{\mathrm{N}}|$
measured using the principle of Fig. \ref{fig:setup}. The detector
is a DC SQUID with two parallel junctions with a maximum critical
current of 3.7 $\mathrm{\mu A}$ and a total capacitance of 100 fF.
Likewise, a JJ with critical current $I_{\mathrm{c,scat}}\simeq 1.6$
$\mathrm{\mu A}$ acts as the noise source. The different curves
correspond to different values of magnetic flux through the SQUID
loop, and therefore to different effective values of
$I_{\mathrm{c}}$ and $\omega_{\mathrm{p}}$ of the detector. For this
particular sample, the quality factor was very low, $Q\sim 2$,
whereby the energy levels are not well separated. A relation between
$I_{0.5}$ and $|\bar{I}_{\mathrm{N}}|$ can be obtained using Eqs.
\equref{ta1} and \equref{prob1} together with $T^{*}$ from
\equref{tstar3}, as discussed in Ref. \onlinecite{pekola05}. With
the factor $QF$ as a fitting parameter, one obtains the solid lines
in Fig. \ref{fig:response} a). Consequently, Fig. \ref{fig:response}
b) shows the variation of these $QF$--values as
$\omega_{\mathrm{p}}$ is altered by changing the magnetic flux. The
measured $QF$--values are generally of the order of 0.01--0.1,
showing a suppression of almost two orders of magnitude compared to
typical values in a similar circuit without special
filtering.\cite{pekola05} This measurement shows that a
SQUID-detector can be used for noise spectroscopy of the circuit and
the noise source. Results of the present measurements are, however,
not yet well characterized. Improved future designs and detectors
with higher quality factors should allow to apply the adiabatic
model.

\section{DISCUSSION AND CONCLUSIONS}
The rather strict limits imposed on the adiabatic model can be
relaxed by using a more general formalism. To treat the
nonequilibrium fluctuations in the setup of Fig. \ref{fig:effective}
without necessarily requiring adiabaticity, one could approach the
problem as quantum tunneling in real time.\cite{schon} This
corresponds to describing the electrical environment and the
fluctuations using an ensemble averaged influence functional and
determining the time-evolution of the reduced density matrix of the
detector junction. Effects arising from the third and higher moments
of fluctuations can then be analyzed by expanding the influence
functional in terms of the cumulants of current. On the other hand,
the adiabatic or resonant excitation models considered in this work
do not take into account the back-action of the detector on the
noise source: Instead, the nonequilibrium current fluctuations are
considered as if they were caused by an external, independent
driving force. This problem can be approached by describing the shot
noise source and the detector as a coupled quantum mechanical system
in terms of an effective action, which can further be related to the
rate of macroscopic quantum tunneling.\cite{nazarov04,ankerhold05}

To summarize, we have analyzed underdamped Josephson junctions as
detectors of current noise. High-frequency fluctuations generated by
a shot noise source are described by a thermal activation model with
an elevated effective temperature. If the junction stays in its
ground state in the absence of transitions induced by high-frequency
noise, the rate of macroscopic quantum tunneling is sensitive to the
higher moments of fluctuations, too. We have discussed the average
escape rate for adiabatic fluctuations which can be directly
compared with measurements. The considered experimental scheme
allows the detection of non-zero higher odd moments using standard
switching measurements provided the high frequencies are efficiently
filtered. Relating the measured tunneling rate asymmetries back to
the properties of the noise source is a particularly important task,
as this involves an accurate characterization of the frequency
dependence of the electrical circuit attached to the detector
junction.

\section*{ACKNOWLEDGMENTS}
We thank J. Ankerhold, O. Buisson, D. Esteve, H. Grabert, T.
Heikkilä, F. Hekking, T. Ojanen, H. Pothier, F. Taddei, and J.
Vartiainen for useful discussions and O-P. Saira for help in the
design of the band-stop filter and calculation of its frequency
response.


\begin{thebibliography}{99}
\bibitem{cumulantnote}  Higher cumulants can in principle be taken
into account in a similar manner, resulting in an asymptotic
expansion around a Gaussian distribution. Furthermore, the
restriction on $\gamma$ can be avoided by direct numerical inversion
of the characteristic function.
\end{thebibliography}
\end{document}